# Artificial intelligence in medicine and healthcare: a review and classification of current and near-future applications and their ethical and social Impact.

**Emilio Gómez-González**[1,2,3], **Emilia Gomez**[4,5], **Javier Márquez-Rivas**[6,7,2,1], **Manuel Guerrero-Claro**[1], **Isabel Fernández-Lizaranzu**[8,1], **María Isabel Relimpio-López**[9,10], **Manuel E. Dorado**[11], **María José Mayorga-Buiza**[12,2,1], **Guillermo Izquierdo-Ayuso**[13,14], **Luis Capitán-Morales**[15].

[1]Group of Interdisciplinary Physics, Department of Applied Physics III, ETSI Engineering School, Universidad de Sevilla. Camino de los Descubrimientos sn, 41092 Seville, Spain.

[2]Group of Applied Neuroscience, Institute of Biomedicine of Seville (IBIS). Avda. Manuel Siurot sn, 41013 Seville, Spain.

[3]Royal Academy of Medicine and Surgery of Seville. C/. Abades 10, 41004 Seville, Spain.

[4]Joint Research Center, European Commission*. C/. Inca Garcilaso 3, 41092 Seville, Spain.

[5]Universitat Pompeu Fabra. Plaza de la Mercé 10, 08002 Barcelona, Spain.

[6]Service of Neurosurgery, Hospital Universitario "Virgen del Rocio". Avda. Manuel Siurot sn, 41013 Seville, Spain.

[7]Center for Advanced Neurology (CNA). Avda. Manuel Siurot 43A, 41013 Seville, Spain.

[8]SeLIVER Group, Institute of Biomedicine of Seville (IBIS). Avda. Manuel Siurot sn, 41013 Seville, Spain.

[9]Department of Ophthalmology, Hospital Universitario "Virgen Macarena". C/. Dr. Fedriani 3, 41009 Seville, Spain.

[10]RETICS, OftaRed, Instituto de Salud Carlos III (ISCIII). Avda. Monforte de Lemos 5, 28029 Madrid, Spain.

[11]Department of Human Anatomy and Embryology, College of Medicine, Universidad de Sevilla. Avda. Sánchez Pizjuán sn, 41009 Seville, Spain.

[12]Service of Anesthesiology, Hospital Universitario "Virgen del Rocio". Avda. Manuel Siurot sn, 41013 Seville, Spain.

[13]Service of Neurology, Hospital "VITHAS-NISA Sevilla". Avda. Plácido Fernández Viagas sn, 41950 Castilleja de la Cuesta, Seville, Spain.

[14]Service of Neurology, DINAC Foundation. C/. Fernando Álvarez de Toledo 4 2º-3, 41009 Seville, Spain.

[15]Department of Surgery, College of Medicine, Universidad de Sevilla. Avda. Sánchez Pizjuán sn, 41009 Seville, Spain.

**Corresponding author:**

**Prof.Dr. Emilio Gómez-González** {egomez@us.es}

Group of Interdisciplinary Physics, Department of Applied Physics III
ETSI Engineering School - Universidad de Sevilla
Camino de los Descubrimientos sn
41092 Sevilla, Spain

www.etsi.us.es/gfi

**Abstract**

This paper provides an overview of the current and near-future applications of Artificial Intelligence (AI) in Medicine and Health Care and presents a classification according to their ethical and societal aspects, potential benefits and pitfalls, and issues that can be considered controversial and are not deeply discussed in the literature.

This work is based on an analysis of the state of the art of research and technology, including existing software, personal monitoring devices, genetic tests and editing tools, personalized digital models, online platforms, augmented reality devices, and surgical and companion robotics.

Motivated by our review, we present and describe the notion of "extended personalized medicine", we then review existing applications of AI in medicine and healthcare and explore the public perception of medical AI systems, and how they show, simultaneously, extraordinary opportunities and drawbacks that even question fundamental medical concepts. Many of these topics coincide with urgent priorities recently defined by the World Health Organization for the coming decade. In addition, we study the transformations of the roles of doctors and patients in an age of ubiquitous information, identify the risk of a division of Medicine into "fake-based", "patient-generated", and "scientifically tailored", and draw the attention of some aspects that need further thorough analysis and public debate.

**Introduction**

Artificial Intelligence (AI) is a new realm of science and technology. It already affects many human activities at all societal levels, from individuals to social groups, corporations and nations. AI is expanding rapidly, worldwide, in almost every industrial, economical and societal sector, from information technologies to commerce, manufacturing, space, remote sensing, security and defense, transport and vehicles and, since the beginning of the XXI century, it is effectively entering into Medicine and Health Care[1,2].

The fast and powerful evolution of AI is fostered by many factors. Among them,

- the availability of powerful and cost-effective computing (processing) tools, hardware (e.g. graphics processing units), software and applications, –even in consumer-grade personal computers and mobile devices– and of large (big) data sets, with many different types and formats of information, both in online and cloud platforms and generated in real time by user wearables and the internet-of-things (IoT),
- the expansion of open source coding resources and online communities of users and practitioners sharing resources, expertise (know-how), and experience, and
- the combination of computer processing with other technologies such as photonics (merging of applied optics and electronics) and human-machine interfaces.

Recent advances in AI systems in Medicine and Health Care present extraordinary opportunities, – particularly in areas of such deep social interest as oncology[3]– together with significant questions[4] and drawbacks[5–7], calling for a close consideration of their implementation[8] and how they affect – and can even change– basic definitions in the medical context[9–11].

This study provides a review of existing and near-future applications of AI in this particular sector from the point of view of their potential benefits and pitfalls, ethical and social impact. We also identify a set of controversial issues that are not deeply discussed in the literature and should be further researched.

Notably, many of the topics presented in this work coincide with six of the thirteen urgent priorities recently defined by the World Health Organization (WHO) for the coming decade[12]. These coinciding priorities explicitly include: "Harnessing new technologies", "Earning public trust", "Protecting people from dangerous products", "Making health care fairer", "Expanding access to medicines", and "Preparing for epidemics". Within the specific priority of "Harnessing new technologies", the WHO defines the challenge as "New technologies are revolutionizing our ability to prevent, diagnose and treat many diseases. Genome editing, synthetic biology and digital health technologies such as artificial intelligence can solve many problems, but also raise new questions and challenges for monitoring and regulation. Without a deeper understanding of their ethical and social implications, these new technologies, which include the capacity to create new organisms, could harm the people they are intended to help".

This work is based on an exhaustive literature review of existing scientific publications and technical publications, including software packages, personal monitoring devices, genetic tests and editing tools, personalized digital models, online platforms, augmented reality devices, and surgical and companion robotics[13]. This review covered 588 references, most of them from 2016-2019 but intended to be dynamically updated, including standard scientific and academic platforms such as MEDLINE, Current Contents and PubMed plus product descriptions and internet and press articles of well-recognized sources which are relevant for the topic. From this review, we summarize here our main findings, discussion points and conclusions, supported by 107 representative references.

**"Tech-savvy": From early pioneers in imaging and surgery to robotics and online platforms.**

In Medicine and Health Care, AI has evolved from computer programs to support the analysis of medical images to its integration in almost every clinical[14] and organizational area[15]. Radiology[16] was at the forefront of this transformation, together with different branches of surgery using augmented reality devices[17] and surgical robots[18]. They were quickly followed by other image-based specialties (e.g. pathology[19], dermatology[20], ophtalmology[21]) and, more recently, by virtually all areas of Medicine and Health Care, from general practitioners[22] to emergency departments[23], epidemiology[24], and disease management[25]. Systems for "computer-aided diagnosis" have expanded to include online assistants (e.g. app, chatbots) –both for very specific medical areas (e.g. oncology[26], predicting the response to treatments[27]) and for the general public[28]–, and "clinical robots" now include "social companions" for hospitalized person, particularly children[29] and the elderly[30]. In addition, wearables and IoT devices allow for real-time monitoring of physiological information, even at home[31], and, integrated with medical and social-media data, can trigger clinically related automated interventions (from suicide prevention calls to police[32] to medication delivery[33]).

Concerned –even healthy– citizens can now order direct-to-consumer genetic tests among many thousands in the market[34]. New tools for big data modeling, analysis, and visualization are also expanding, and provide substantial, transforming improvements in clinical pathways[35], from the generation of "digital twins"[36] of individual patients to self-management of treatments[37]. There are even online, crowd shared platforms for such high-end applications as radiotherapy[38]. Many management aspects related to health economy (e.g. increased efficiency, quality control, fraud reduction) and policy[39] also benefit from the new AI mediated tools. They even offer new hopes of improvements in health for environments with reduced resources[40] and in developing regions[41].

However, technical challenges[42–44] and ethical concerns[45–47] remain, and new important questions arise.

**Classification of the applications of Artificial Intelligence in Medicine and Health Care.**

In Figure 1 we propose a graded classification of AI and AI-mediated applications in Medicine and Health Care according to their beneficial vs detrimental character as recognized in the reviewed literature. For each of these applications, we report the technology, the specific implementations behind and their level of availability according to published references. On the one side of the spectrum of social impact we consider applications that have been shown to be beneficial to support clinical decision making, while on the other side we place applications that are widely agreed to be harmful.

In Figure 1 we also propose a (new) scale named "Technology Availability Level" (TAL) which gives a qualitative description of the degree of availability of a technology, in a numerical scale from 0 (unknown) to 9 (available for the general public). The TAL scale is similar in format (and related) to the standard "Technology Readiness Levels" (TRL) but, as mentioned, it is based on published references (in scientific and academic literature, industrial or corporate reports, and general media citing sources considered to be reliable according to standards). It is important to consider that "availability" is not necessarily equivalent to "readiness levels" due to such factors as disclosure according to industrial, proprietary and/or government strategies, and that the TAL scale does not evaluate either the fulfillment of regulatory processes. The values defined for the TAL scale are the following:

TAL 0. Unknown status. Not considered feasible according to references.
TAL 1. Unknown status. Considered feasible according to related, indirect references.
TAL 2. General/basic idea publicly proposed.
TAL 3. Calls for public funding of R&D open.
TAL 4. Results of academic/partial projects disclosed.
TAL 5. Early design of product disclosed.
TAL 6. Operational prototype/'first case' disclosed.
TAL 7. Products disclosed but not available.
TAL 8. Available for restricted (e.g. professional) users.
TAL 9. Available for the public.

| AI and AI-mediated technologies | Specific implementations. | TAL | Social Impact |
|---|---|---|---|
| Algorithms for computer-aided diagnosis. | SW for decision support in (most) clinical areas. | 8, 9 | Positive |
| Structured reports, eHealth. | SW for improved workflow, efficiency. | 8, 9 | |
| AR/VR, advanced imaging tools. | Tools for information visualization and navigation. | 6, 7, 9 | |
| | Image-guided surgery. Teleoperation. | 4, 6, 9 | |
| Digital pathology, 'virtopsy'. | SW for automated, extensive analysis. | 4-9 | |
| Personalized, precision medicine. | Tailored treatments. Prediction of response. | 4-9 | |
| | 'In-silico' modeling and testing. The 'digital twin'. | 4-8 | |
| | Drug design. | 4, 8 | |
| Apps, chatbots, dashboards, online platforms. | The 'digital doctor' (assistance for professionals and for patients). | 8, 9 | |
| Companion and social robots. | For hospitalized persons, children & the elderly. | 4-9 | |
| Big Data collection and analysis. | Epidemiology, prevention and monitoring of disease outbreaks. | 2-9 | |
| | Fraud detection. Quality control, monitoring of physicians and treatments. | 4-9 | |
| IoT, wearables, mHealth. | Automated clinical/health surveillance in any environment/institution. | 7, 8 | |
| | Monitoring, automated drug delivery. | 7-9 | |
| Gene editing. | Disease treatment, prevention. | 7, 8 | |
| Merging of medical and social data. 'Social' engineering. | Prevention of episodes with clinical relevance (e.g. suicide attempts). | 6, 8 | Controversial |
| | Tailored marketing (e.g. related to female cycles). | 6, 8 | |
| Reading and decoding brain signals. Interaction with neural processes. | Treatment of diseases. Restoring damaged functions. | 3-8 | |
| | Brain-machine inferfaces. | 5-8 | |
| | Control of prostheses, exoskeletons. 'Cyborgs'. | 2-7 | |
| | Neurostimulation. Neuromodulation. | 4-8 | |
| | Neuroprostheses (for the central nervous system). | 2-5 | |
| | Mind 'reading' and 'manipulation'. | 1-3 | |
| Genetic tests. Population screening. | Disease tests. Direct-to-consumer tests. | 4-9 | |
| Personalized, precision medicine. | Individual profiling. Personalized molecules (for treatment) at 'impossible' prices. | 3-8 | |
| Gene editing. | 'Engineered' humans. | 2, 6 | |
| | Gene-enhanced 'superhumans'. | 2 | |
| | Self-experimentation medicine. Biohacking. | 2, 6 | |
| Fully autonomous AI systems. | The 'digital doctor'. | 2-5 | |
| | 'Robotic surgeon'. | 2, 4 | |
| Human-animal embryos. | Organs for transplants. | 2, 4, 5 | |
| | Hybrid beings ('chimera'). | 2, 4 | |
| The quest for immortality. | Whole-brain emulation / 'transplant'. | 1, 2 | |
| The search for artificial life forms. | 'Living machines' ('biological robots', 'biobots') | 4, 6 | |
| | Military. | 2, 3 | |
| Evil biohacking. | Targeting specific individuals or groups. | 1, 2 | |
| Weaponization. | From 'small labs' to military labs. | 1, 2 | |
| Bioterrorism. | From 'small labs'. | 1, 2 | Negative |

Figure 1. Proposal of classification of artificial intelligence (AI) and AI-mediated applications in Medicine and Health Care according to their beneficial vs detrimental character. SW = software, AR = augmented reality, VR = virtual reality, IoT = internet of things. TAL = Technology Availability Level (see details of values in the text).

In Figure 2 we complement this classification with an analysis of the main related (ethical and social) aspects of AI technologies by defining three partially overlapping sets:

1. In the first group we include those topics commonly under analysis as raised by other AI application domains (e.g. social networks, online commerce, automation in factories, autonomous vehicles) such as data use and sharing[48,49], privacy[50], security, and annonymization[51].

2. In the second group, we consider topics –some of which are also under analysis in other sectors– which are of particular relevance in our context. Among them, the proper training of AI systems to avoid bias[52] and assure fairness and equity[53], the difficulties to explain the "reasoning process" of complex systems[54], the lack of updated evaluation protocols[55,56] and regulatory standards[57,58], the legal aspects of liability and malfunction[59], and the vulnerability of medical information under adversarial attacks[60]. And, of course, the question of whether (or not) harnessing AI systems under human control[61].

3. Finally, in the third group we place some aspects barely or not included in the reviewed literature. They mainly relate to the potential dual use of certain technologies, from limit-defying research about human control, 'biological robots', human-animals hybrids and biohacking to weaponization and bioterrorism. They will be discussed –with examples from neuroscience and gene editing– in a later section.

It is important to note that many AI systems and AI-mediated applications show an intrinsic "mix" of positive, negative, and controversial aspects depending on their specific implementations, and that, according to published information, their readiness levels vary from commercially available to very early, conceptual designs.

To our knowledge, this is the first attempt to provide an exhaustive list of AI systems and applications in Medicine and Health Care taking into account their potential benefits and pitfalls. However, it is not intended to define an "absolute" scale of "goodness" or "badness", as many technologies (e.g. gene editing, neuroprostheses) are not necessarily "positive" or "negative", and others may certainly be difficult to categorize.

Even the scientific and ethical criteria for such analysis need a review, and new questions arise. The methods used to evaluate the performance of medical products and treatments are based on averages over population groups, while we now consider extremely personalized approaches, to the genetic structure of each individual. How can those treatments be rigorously tested? Which are the time and cost required to find "enough cases" to "generate scientific evidence"? In addition, how should AI systems be benchmarked? Should they be compared to a (possibly error-prone) human doctor or "against" another "machine"? Common ethical guidelines for the evaluation of technologies mostly date from the pre-digital era. Which should be the figures of merit to consider? Which are the roles of the public and the policy makers?

The goal of the classification proposed in these figures is to draw attention to the need of a careful, thorough analysis and as the basis for our discussion below, with a focus on those issues hardly addressed in the current literature. As mentioned, it is particularly important that many of these topics have been explicitly defined as "urgent priorities for the coming decade" by the WHO[12] at the beginning of 2020.

| (G1) Currently under analysis, as raised by other areas of AI applications. | |
|---|---|
| **Aspects.** | **Questions.** |
| Data privacy, integrity. | Ownership. Authorization for data collection, sharing, mining, exchange. |
| Anonymity. | Surveillance anxiety. |
| Responsibility. Accountability. | Who is responsible in case of malfunction? |
| Effects on professionals and employment. | Lost & new jobs. Deep changes in some medical specialties (some may even disappear). Need of professional updating. Quality control, monitoring. |
| Security. Reliability. | Vulnerabilities. Data theft. Manipulation of the data used to train the systems. |
| Performance. | Improved health outcomes and clinical pathways. Reduction of medical errors. 'Personalized Medicine'. Psycho-social outcomes. |
| Human-in-the-loop? | Should a human operator override AI systems? Even if human is more 'error-prone'? What happens if there is no time to act? |

| **Aspects.** | **Controversies.** | |
|---|---|---|
| Explainability. | Currently required by legislation. Some systems are (will be) too complex to be understood by a human. But they may give better results than a human. | **(G2) Of particular relevance for AI applications in Medicine and Health Care.** |
| Trust. | Does 'the machine' perform better than a human doctor? What to do if they (AI system, human doctor) give conflicting opinions? 'Digital health scammers'. | |
| Data quality. Bias / fairness. | Do AI systems have biases/are fair with different (e.g. ethnic, gender, age) groups? Do they receive proper, balanced data for training? Are results valid? | |
| Empathy. | Shared decisions? Help (the human) take difficult decisions? | |
| Citizen (taxpayer) opinion and involvement. | Common-good in public-funded research, informed consent, citizen science. Reduced 'asymmetry' doctor-patient. 'Patient-centric' model. | |
| Test, benchmarking. | How to evaluate results? Existing procedures for average groups are valid for individualized treatments? Comparison of AI systems 'against humans or machines'? | |
| Regulation. | Lags behind technology. No international consensus. | |
| Affordability. Economic impact. | Optimal treatments at 'impossible' prices? A factor of inequality? New models for health insurance and coverage? | |
| Information for the public and professionals. | Pressure for new products. Real advances vs hypes and non-confirmed stories of success in areas of great interest (e.g. cancer cures). Risk of 'fake-based' medicine. | |
| Life and death decisions. | Should we allow 'a machine' to take them (on us, on a relative)? The debate about lethal autonomous weapon systems. | |

| **Aspects.** | **Significant/conflicting issues.** | |
|---|---|---|
| Humanization of care. | Professionals with AI: More time with the patient, stress relief. AI systems: Currently, lack of physical exam/contact with patient. | **(G3) Barely/not included in analysis of AI applications in Medicine and Heath Care.** |
| Social engineering, profiling based on merged medical, health, social data. | Preventive detection of events (e.g. suicide) vs tailored marketing, insurance, health care, employment. Genetic screening of the population. | |
| Availability of (unsupervised, unreliable) multiple data, genetic tests for anyone. | Risk of 'patient-generated' medicine. | |
| Limits to data use? Post-mortem, inheritance. | Post-mortem use of individual (e.g. genetic) information? | |
| Crowd-sourcing of algorithms, processing power. | Free sharing of expertise, know-how, experience. Solidarity vs risks of malicious use. | |
| Reading, decoding brain signals. | Hope for severely impaired vs privacy at its basics. | |
| Interaction with neural processes. | Help for neurological, mental diseases vs free will. | |
| Gene editing as self-experimentation. | Risk of unexpected results. Change of genetic heritage. | |
| Gene editing of (human, human-animal) embryos. | Risk of unexpected results in newborns. Creation of new beings ('chimera'). | |
| The two sides of technology. | 'Easy' weaponization. High risk for bioterrorism. | |
| Whole-brain emulation / 'transplant'. | The quest for immortality. Definition of life. | |
| 'Living machines' ('biological robots', 'biobots') The search for artificial life forms. | Definitions of life (natural, artificial) and death. | |
| **Benefits versus risks and pitfalls.** | **Limits (or no) to research and development?** | |

Figure 2. Ethical and social aspects of artificial intelligence (AI) and AI-mediated technologies in Medicine and Health Care in three groups (G1, G2, and G3). Some key relevant questions, controversies, significant, and conflicting issues are outlined for each aspect.

**"Extended Personalized Medicine".**

The original goal of Personalized Medicine is to exploit very specific biological (genetic) features of individuals for tailored diagnosis and treatment[62]. Decoding the genome of each patient represents a very significant change from the existing model of averaged analysis of populations to an extremely individualized approach[63], for treating disease –in a new paradigm defined as "precision medicine"[64]– but also to promote wellness[65] and healthy, personalized lifestyles.

However, although not explicitly formulated in the literature, we consider that the underlying principle of Personalized Medicine can be further expanded. It can include other properties whose particular values or structures –even their spatial distribution and time evolution in the human body– may be significantly different for any single individual, in different clinical situations, at every moment of life and, possibly in strong relationship to each other. Such additional features form the new concept of "*Extended Personalized Medicine*" and they may come from

- "known sources" from the "basic sciences" of physics (e.g. bioelectromagnetic fields and signals, biomechanical magnitudes and properties, hydrodynamic parameters of the circulation of any fluid in the body), chemistry (concentrations of ions, molecules), and biology (metabolites),

- "not yet known" origins. This concept refers to the potential characterization of brain processing schemes, connections, and functions whose details still remain veiled for science,

- demographic data, extracted from conventional databases,

- data about psychological and emotional status, extracted indirectly from the individual activity in social networks.

- social data, including those about societal structures (family, groups providing psychological, emotional support), and cultural and religious beliefs which may influence health-related issues, such as restrictions on types of food or sexual activity, provided by the user or mined from social networks,

- "lifestyle parameters" (sleep hours, stress, physical activity, food ingest) easily accessible through apps, wearables, and IoT,

- values of environmental and physical geography conditions (weather, contamination) transmitted by multiple platforms,

- sensors evaluating mood through face and gesture recognition, changes in cardiac rhythms, perspiration, and breathing patterns when receiving certain visual or auditory stimuli. They may be biometrics readers in smartphones, domotic environments, and wearables,

Many sources of such data are already available. They include "standard" clinical scores and evaluations (imaging scans, analytics), results of genetic and "omics" (genomics, metabolomics, proteomics) tests, and medical knowledge from publications and references, but also data from social networks[66] and from personal, IoT, and home sensors[67]. Together with potential benefits[68], such as predicting individual response to treatments[69], this "accumulation" of personal, intimate information also presents very high risks regarding ownership, security and privacy.

From a technical point of view, the combination of varied heterogeneous sources requires the use of advanced AI-mediated tools for merging, processing, and analyzing multiple data layers – extracting useful information– and to operate the devices of augmented and virtual reality for the (very much needed) interactive navigation and visualization.

**Replacement or enhancement: Do we need a doctor?**

AI systems pose important conceptual challenges for physicians. The most relevant is the debate between *AI-replaced* or *AI-enhanced* doctors[70,71].

As in any other sector, the appearance of (even partially) automated systems produces deep professional transformations, as some jobs will be lost and new ones will arise. Current systems for clinical applications may be mostly considered as "weak AI" that is, they can succeed at "narrow" tasks and be useful helping human physicians in tedious, repetitive activities (e.g. finding elements or segmenting organs or structures in images). As mentioned, medical specialties dealing with visual analysis of images have started to experiment substantial changes through such "aids". Bottom-line is that computerized systems can process many, many more images than any human operator can, without fatigue and time constraints, and –if properly trained– with even better results. AI systems also show very good performance in the recognition of natural language and written texts, which in turn opens the way to extracting information –for systems to learn– from many available sources (e.g. clinical records) which are not usually available in structured, useful formats. But in general, medical AI systems lack the ability to "interpret" the context and "generate" the most distinctive human features (creativity, emotions), and cannot be considered (yet) as "strong" or "general" AI[72]. Although many questions remain[73], AI systems can enhance many aspects of doctor's tasks and those professionals choosing not to use them will be outdated.

The availability of massive amounts of data conveying very different types of information of each individual patient also requires new professional profiles –not necessarily physicians– to explore and extract the most useful items for diagnosis and treatment. Genetic counselors are already joining clinical teams to help (patients but also professionals) in understanding the complex gene-related information[74]. The addition of novel sources of data –in the foreseeable paradigm of "expanded personalized medicine"– will likely require some other profiles of "medical data scientist" for specialized advice for physicians and patients.

**Key intangible qualities.**

Healthcare is not just an intellectual interaction but highly social, especially relevant in situations of vital relevance (e.g. decisions with effects on life quality, palliative care, euthanasia). There are many different views on trust, empathy, and the humanization of care, and how they can be affected by the new reality of automated systems, from the reduction of medical errors to the possibility of AI-systems helping humans to take difficult decisions, to the design of better doctor-AI and patient-AI interaction strategies[75,76]. Proponents of AI-enhanced human doctors see the integration of new tools as a powerful way to regain human aspects of the relationship between the doctor and the patient[77] but the real results are yet unknown[78].

A key issue in of the greatest importance in all AI-related areas is that of increasingly autonomous systems[79]. In the medical and surgical fields, their capabilities may include the decision about human lives and, if they ever become available, which would be their ethical guidelines? For many people, this scenario may seem unreal, but there is a current debate about a topic that shares some of the arguments and positions, namely that about lethal autonomous weapons systems. Such devices have already made their way to the public –even to be discussed at the United Nations[80] – as this type of research has been (at least) partially disclosed and early systems begin to be displayed. Their objectives are clearly the opposite of medical devices, and the popular name of "killer robots" may ward it off from being included in medical literature, but the fundamental idea to discuss is the same: will "a machine" take the ultimate decision to keep or end a human life?

**A new situation: patients take the lead.**

The evolution of individual behavior as related to Medicine and Health Care presents a novel array of many advantages, pitfalls, and un-addressed concerns. The overall access to many types of data has already had an important effect in the relationship between the patient and the doctor, namely the reduction of the "asymmetry in information" between them and the evolution towards a "patient-centric" model[81]. This new situation started with the generalized availability of information in online platforms. Suddenly, patients could ask "the Internet Doctor" about anything[82], from symptoms to the side effects of treatments to advices for healthy lifestyle, and then visit the real physician's office with a list of "informed" questions, requests, and even complaints. In the following years, it has become evident that there is no "a priory" guarantee of the quality –even the certainty– of the information found on internet searches. Very valuable resources are mixed with completely erroneous –even maliciously misleading– material and a certain level of knowledge is required to find and understand the information of real interest for any case. To evaluate the clinical situation of a patient and potential treatment options there is a clear need of the "integrated analysis", of the "global vision" provided by a qualified, trained, real doctor.

The evolution of technology is expanding AI systems, starting from "basic" –but very effective– symptom's checkers and on the road to increasingly autonomous "digital doctors". AI-supported, even shared-decisions –with non-human systems– and patient involvement shape substantial changes. However, a very dangerous division of "Medicine" in three sub-types may therefore take place:

i) *"Fake-based" Medicine*. Based on (unfounded, unconfirmed) rumors and "fake news", this type of "pseudo-medicine" may present "ancient, natural knowledge" as opposed to scientific, evidence-based medicine, considered to be under malicious control by corporations, academia, institutions, and governments. Even rejecting technology, it may easily take benefit from the expanding ability and multiplying power of online platforms and AI-mediated tools (including chatbots, interactive apps, and communities of followers) for dissemination of wrongful information. This type of misinformation –such as in the case of "anti-vaccine" groups– is increasing[83], being used to discredit "conventional" therapeutic approaches and to promote that patients abandon treatments and follow-up by physicians, with very serious potential consequences –even with the risk of death– both for the individuals affected and their surrounding environments. In addition, *"digital health scammers"* can benefit from available data-driven personalized AI tools to offer clinical advices and treatment options to vulnerable population with no guarantee or reliability.

ii) *"Patient-generated" Medicine* deriving from the growing online availability of many (both correct and unsupervised, unreliable) sources of medical information, even of platforms and apps supposed to evaluate and interpret the results of (almost any) type of analysis, including imaging scans and genetic tests.

Although a "better informed patient" is a positive consequence of the availability of information, individual-ordered analysis and diagnosis lack the (fundamental) "global vision" that the doctor can offer to the patient and the (crucial) trained skills required for proper understanding of test results and deciding subsequent steps. Any person, even medically illiterate, without any medical education or training, may have –through AI-mediated tools– immediate, unlimited access to a trove of information that may be considered correct and related to a disease or health issue. Resulting decisions may then –very probably– bring inadequate, even damaging, consequences, without the potential help or support from any established medical institution.

iii) *"Scientifically tailored" Medicine* is the type of medical science that evolves from research into extended personalized medicine. For the patients, the critical decision would probably be the selection of the human doctor –perhaps the AI-system– to lead the team of "conventional" (clinical) and "new" (e.g. genetic counselors, medical data scientists) professional profiles required to correctly integrate the multiple, extensive sources of information to establish the diagnosis and define the corresponding treatment and monitoring strategies.

**Affordability and inequality.**

Global figures and market of AI in Medicine and Health Care forecast very relevant, positive impact for the coming years. However, the economic analysis must include the ethical and social points related to health systems, the industries and the patients[84]. It is important to note that the cost of decoding a human genome is substantially low –in the order of a few hundred euros– but the prices of some of AI-mediated treatments, such as certain personalized drugs, may reach "impossible" figures, even in the order of millions of euros per case. This steep step is due to the difficulties of individually tailoring drug molecules to the specific genome of an individual. New models of health coverage, insurance, and affordability may be needed[85] as such clinically excellent technologies pose a clear risk of evolving into a significant increasing factor of inequality for most people.

**Barely mentioned but not science fiction any more: the dual use of technology.**

We address now some aspects and applications of AI-mediated technologies that can be very controversial and that, even though society is not currently addressing, are advancing at a fast, uncontrolled, unsupervised pace. Most developments simultaneously present the potential of very positive, disruptive improvements with deeply disturbing, ethically questionable –even very negative– outcomes.

i) Decoding and interacting with brain signals: from hope to limitations on free will.

Neuroscience is one of the scientific realms in which the advent of technological advances from other disciplines is fostering an extraordinary development. Main contributions come from merging AI, photonics and engineering. Neuroscience also encompasses neurosurgery and neurology with other clinical areas (pharmacology, psychology) and related sciences (biology and genetics, biochemistry). With significant ethical challenges[86], advances point towards progressively non-invasive, remote reading and decoding the complex signals of the brain and the design of advanced man-machine interfaces[87], with simultaneous health-related[88] and military applications[89].

On an extraordinarily beneficial approach, this knowledge opens the way to the human nervous system –already starting to allow for interactive control of innovative, active prostheses–, which offer a great hope for many persons with severely disabling conditions, and many potential industrial applications. There are very strongly funded projects, e.g. the European Human Brain Project[90] and the USA Brain Initiative[91], oriented to "the positive aspects" of neuroscience. However, this knowledge also relates to very controversial paths, e.g. the potential ability to "read the mind", and its eventual combination with different types of neural stimulation and interaction with brain signals, which, in turn, might lead to undesired forms of manipulation (lack of free will) and human control.

ii) Gene editing: From tailored treatments to unexpected effects, self-experimentation, human-animal embryos, biohacking and bioterrorism.

A clear example of a controversial, AI-mediated area is human gene editing. The possibility of altering –and substituting– definite fragments of the genetic (deoxyribonucleic acid, DNA) chain of human cells is a challenging task –which requires AI tools to reach the required precision– with many potential applications.

Genetically modified organisms are becoming common in agriculture and farming to increase productivity and reduce the impact of plagues and diseases, and the subject of relatively active social debates in some countries. In 2018, human gene editing –and the ethical doubts– made to the headlines as the first two "engineered twins" were born in China[92]. From a "positive approach", the "correction" of "wrong", mutated genes to avoid severe diseases, important questions immediately arise: Does it really work? In terms of impact on the human genetic heritage, at what cost? It has already been published that these two babies present unknown defects and unpredictable future development. Then, is it ethical to design "experimental human beings"? What will happen if those people become adults and procreate? Will there be any limits in the genes to be modified? That is, if the procedure is technically feasible, will anyone engineer some types of "super-humans"? Should there be some type of oversight of this research? Most of these questions are common to other controversial areas of AI applications, have barely been asked and do not have any clear answer yet.

Another troublesome application of gene editing technology is the generation of human-animal embryos. They are purportedly created with the goal of solving the critical scarcity of organs for transplant and under the self-imposed limits of their creators of developing only required organs in genetically modified, not viable, embryos or not letting new beings grow beyond certain thresholds. Perhaps this type of research might even lead to growing "replacement organs" without the need of a "hosting embryo". But if some types of them might eventually have the potential to develop into "adult beings", should they be allowed? Would they be considered "full or partially" humans subject to rights? In ancient mythology, these types of creatures were called "chimera" and their existence was limited to the godly heavens and the infra-world of hell. But in 2017, early experiments of interspecies (human-pig) cell growing were announced, in 2018 first steps in a human-sheep embryo was disclosed, and in 2019 Japan formally authorized this type of research under certain restrictions[93]. A ratcheting up in this field is given by the recently disclosed development of a prototype of a 'living machine' –also considered to be a 'biological robot' ('biobot')– based on animal cells[94], and the search for artificial forms of life for military applications[95].

With a relatively small laboratory and the proper knowledge and tools, gene editing can be achieved. There are some reports of persons undergoing self-experimentation, who modify their DNA and re-introduce it in their bodies, in a novel type of "citizen science approach" named *biohacking*[96]. Their intended purpose is generally to obtain "enhanced capabilities" but this type of research also presents a potential interest for the health market[97]. However, the availability of certain AI-mediated tools (hardware, software, and datasets) may also open the way to malicious forms of biohacking. Gene editing combined with such use of AI technologies present a worrying, very soft boundary to weaponization and bioterrorism[98]. Genetically modified organisms can be very difficult to detect and track. They might be even stored and transported as altered cells in a carrier person targeting specific individuals or populations.

**Informed public debate on setting (or no) limits to research and development.**

AI in Medicine and Health Care is no longer an isolated discipline but laying at the intersection of medicine, sciences and engineering, and social disciplines (ethics, philosophy). It already has deep influences on many issues related to human health and wellbeing at many different scales and the consequences of the wide range of ongoing developments may affect basic concepts covered by international regulations, from the Hippocratic Oath in Medicine to the fundamentals of free will as declared in the Bill of Human Rights.

In addition, AI in Medicine and Health Care is a very dynamic area, inevitably linked with (the very fast) evolution of other technologies, mainly those related to computers, photonics and different branches of engineering. It has many positives perspectives together with controversial and clearly negative aspects. In some applications, AI and AI-mediated tools even evolve to questioning the definition and the limits of life. But which is the public view about such research[99,100]? Most scientific and technological initiatives receive public funds and society needs informed decisions and rigorous information to general and professional audiences, as in other areas of AI[101–103]. Some specific guidelines begin to appear for big data and digital health[104,105], and even a "do no harm" approach has been formulated recently[106], but most of the applications of AI in Medicine and Health Care shown in Figure 1 –namely, those that present the most conflicting aspects– are neither mentioned nor being analyzed yet. As mentioned, some of these topics have recently defined as urgent priorities for the coming decade by the WHO[12].

However, the required study is not easy. The categorization of technologies as "good" or "bad" is strongly dependent on many factors, in which subjective, personal and societal values play an essential role. While it seems evident that objective, professional criteria should be defined for proper judgement, the extraordinary impact of Medicine and Health Care in human beings and society transcends a "technical evaluation" and requires to take into consideration cultural, ethical and social criteria as well. To increase the complexity of the analysis, the acceptance and potential expansion of the applications of AI relate to multifaceted balances of welfare effects and economic, geographical –even political– aspects, but they are essentially linked to the personal, intangible qualities –based on trust– that configure the (current) relationship of doctor and patient.

At the "person scale", important goals of public, regulatory institutions should be to allow for the (very challenging) generalized access of the population to the beneficial results of "scientifically tailored" Medicine while protecting citizens from falling into the risks of "pseudo-medicine" ("fake-based" and "patient-generated") and into becoming victims of "digital health scammers". To achieve such objectives, the availability of fair, trustworthy, contrasted information open to public access is essential.

Moreover, at the "institutional scales", we propose to follow the example of another "controversial" research area, namely that of nuclear technology, which fosters beneficial applications –as in radiation therapy against cancer– while simultaneously restricting negative developments –as weapons of massive destruction–. Key aspects of AI and AI-mediated technologies related to the most controversial topics could be identified and have their access and availability restricted only to qualified users, following similar protocols and procedures to those used to control nuclear and radioactive sources, devices, and design specifications. Definite efforts are required to thoroughly study each element included in the proposed classification and the new societal and ethical challenges that they generate.

**Conclusions.**

The range of applications of AI and AI-mediated technologies in Medicine and Health Care is vast and rapidly increasing, with many powerful potential (positive and negative) results, which may affect the human being and society at all scales. Most of the questions collected in this review remain challenging as their answers are not yet clear at this time, but our goal is to open the way for a multidisciplinary, public discussion of the raised issues to define the principles, ethical and societal guidelines –and potential boundaries– on this matter.

**Acknowledgements.**

The study was conceived, developed, and executed by the authors with no external or commercial support, and was partially funded by the Center for Advanced Studies of the Joint Research Centre of the European Commission through the HUMAINT Project [107].

**Declaration of interests.**

The authors declare no competing interests.

**Contributions.**

Study design: EGG, EG.

Literature search, data collection: MGC, IFL, EGG, EG.

Figures: EGG, EG, JMR.

Data analysis and interpretation: EGG, EG, JMR, MGC, IFL, MIRL, MED, MJMB, GIA, LCM.

Manuscript writing: EGG, EG.

Critical review: EGG, EG, JMR, MGC, IFL, MIRL, MED, MJMB, GIA, LCM.